\documentclass{PoS}
\usepackage{psfrag}

\newcommand{\be}{\begin{equation}}
\newcommand{\ee}{\end{equation}}
\newcommand{\bea}{\begin{eqnarray}}
\newcommand{\eea}{\end{eqnarray}}
\newcommand{\bi}{\begin{itemize}}
\newcommand{\ei}{\end{itemize}}
\newcommand{\non}{\nonumber}
\newcommand{\pho}{\phantom{1}}
\newcommand{\phm}{\phantom{-}}

\title{Automatic generation of vertices for the Schr\"odinger functional}

\ShortTitle{Automatic generation of vertices for the Schr\"odinger functional}

\author{\rightline{\it HU-EP-07/44}}
\author{\rightline{\it SFB-CPP-07-54}}

\author{\speaker{Shinji Takeda}\\
        Humboldt Universitaet zu Berlin\\
        E-mail: \email{takeda@physik.hu-berlin.de}}

\author{Ulli Wolff\\
        Humboldt Universitaet zu Berlin\\
        E-mail: \email{uwolff@physik.hu-berlin.de}}

\abstract{We present a multiplication algorithm to recursively
construct vertices
for the Schr\"odinger functional in the abelian background field case.
The algorithm is suited for automatic perturbative calculations with
a variety of actions.
As first applications,
we derive ratios of the lambda parameters between
the lattice scheme (improved gauge actions including six link loops)
and the $\overline{\rm MS}$ scheme,
and one-loop results
for the Schr\"odinger functional coupling with a lattice $T=L \pm a$,
which is motivated by considering staggered fermions.
}

\FullConference{The XXV International Symposium on Lattice Field Theory\\
		 July 30 - August 4 2007\\
		 Regensburg, Germany}

\begin{document}

\section{Introduction}

It is well known that in lattice gauge theory
the vertices are quite complicated
especially for the pure gluon sector.
This is because, on the lattice,
we are preserving gauge invariance at finite cutoff and
the lattice itself breaks the Lorentz symmetry explicitly.
This is the main difficulty of lattice perturbation theory.
To reduce the risk of errors and to alleviate the  tedious task
of deriving the vertices,
it is desirable to have an automatic method.
A first attempt was made by L\"uscher and Weisz about twenty yeas ago
\cite{Luscher:1985wf}.
They worked in momentum space and
performed some calculations by using their algorithm
which was restricted to closed loops sufficient
for pure gauge theory.

Recently a new algorithm, that we call bottom up algorithm,
was proposed by Hart et. al.\cite{Hart:2004bd}.
A crucial point in this generalization  is that
it can deal with any parallel transporter,
not only with closed loops.
This allows to also include
a fermion action
or even a smeared HQET action.
A relevant assumption for this algorithm has been translation invariance.
Our main concern in this note is
to extend the bottom up algorithm to the Schr\"odinger functional
(SF)\cite{Luscher:1993gh} where
this invariance is broken for the time direction.
Before going to the extension we first summarize the 
position space version of the
algorithm on
the usual translation invariant lattice in the next section.

\section{Bottom up Algorithm}

In Ref.\cite{Hart:2004bd},
the authors explain the algorithm in momentum space
but here we will move to coordinate space.
In the following we denote the antihermitean 
gauge fluctuation field by $q_{\mu}(x)$,
and the link variable (still without background field) by
$U(x,\mu)=\exp (g_0 q_{\mu}(x))$.

A first important point for the automatic operation
is how to represent the vertices in a program.
We consider the parallel transporter along a curve
${\cal L}$ on the lattice.
The $r$'th order coefficient in the Taylor expansion $P_r$ of
the parallel transporter $P[{\cal L};q]$ is written as
\be
P[{\cal L};q]
=
\sum_{r=0}^{\infty}
\frac{g_0^r}{r!}
P_r[{\cal L};q],
\hspace{10mm}
P_r
=
\sum_{a_1,...,a_r}
{\cal C}_{a_1\cdots a_r}\sum_{k=1}^{N_r}
q^{\alpha_k^1}_{a_1}\cdots q^{\alpha_k^r}_{a_r}
f_i,
\label{eqn:P}
\ee
where $\alpha=(\mu,x)$ is a combined index labelling links,
and $a$ is color.
We call ${\cal C}_{a_1\cdots a_r}=T_{a_1}\cdots T_{a_r}$ a color factor,
and $f_k$ an amplitude.
The latter corresponds to a value of a reduced vertex
with index $(\alpha^1_k,\cdots,\alpha^r_k)$.
The reduced vertex here means a vertex without the color factor.
The information about the reduced vertex of order $r$ is now encoded in a list
consisting of $N_r$ lines
\be
L_k^{(r)}=(\alpha_k^{1},\cdots,\alpha_k^{r} ; f_k),
\ee
which holds the index structures and amplitudes
of the corresponding reduced vertex.
The list $L^{(r)} = \{L_k^{(r)}|k=1,..,N_r \}$
corresponds to the all nonzero elements of the reduced vertex of order $r$.
It will be constructed recursively by the multiplication algorithm
to be given below.
Note that in eq.(\ref{eqn:P}) the color factor is located
outside of the $k$-summation,
because it is independent of the shape of the transporter.
However, for the SF this nice structure is rendered more complicated
by the background field as we will see in the next section.

As an explicit example for a list, let us look at
 the case of
the single gauge link variable.
From the expanded form
we can obtain a list $L^{(r)}$ of the link variable,
\be
U(x,\mu)
=
\sum_{r=0}^{\infty}
\frac{g_0^r}{r!}
\sum_{a_1\cdots a_r}
{\cal C}_{a_1 \cdots a_r}
q^{\alpha}_{a_1}\cdots q^{\alpha}_{a_r},
\hspace{8mm}
L^{(r)}=(\underbrace{\alpha,\cdots,\alpha}_{r \mbox{ elements}};1),
\hspace{8mm}
\alpha=(\mu,x).
\ee
It consists of only one line ($N_r=1$)
and is the elementary building block of the algorithm,
that is, something like an initial condition.

So far we have defined the fundamental elements on which the algorithm
operates.
Next
we consider the case that
a parallel transporter $P$ is composed from
$P^{\prime}$ and $P^{\prime\prime}$, $P=P^{\prime}P^{\prime\prime}$
and want to obtain the vertices of $P$ from those of
$P^{\prime}$ and $P^{\prime\prime}$.
In other words,
the problem is how to obtain a set of lists of $P$ (up to a certain order),
from those of $P^{\prime}$ and $P^{\prime\prime}$,
$\{L_i\}_{P^{\prime}} \times \{L_j\}_{P^{\prime\prime}}\longrightarrow \{L_k\}_{P}$.
Since any parallel transporter is composed of elementary one-link variables,
by repeating the procedure,
we can obtain the lists for arbitrary parallel transporters.
This is the origin of the name  `bottom up'.
The algorithm
can be easily understood by looking at an actual
multiplication of the coefficients of the Taylor expansion.
The coefficient of $P$ with order $r$,
$P_r$, is expressed by those of $P^{\prime}$ and $P^{\prime\prime}$ as
\bea
P_r
&=&
\sum_{s=0}^{r}
\frac{r!}{s!(r-s)!}
P_s^{\prime}P_{r-s}^{\prime\prime}
\non\\
&=&
\sum_{a_1...a_r}
{\cal C}_{a_1...a_r}
\sum_{s=0}^{r}
\frac{r!}{s!(r-s)!}
\sum_{i=1}^{N_{s}^{\prime}}
\sum_{j=1}^{N_{r-s}^{\prime\prime}}
q^{\alpha_i^1}_{a_1}... q^{\alpha_i^{s}}_{a_s}
q^{\alpha_j^{1}}_{a_{s+1}}... q^{\alpha_j^{r-s}}_{a_r}
f_i^{\prime}f_j^{\prime\prime}
\non\\
&=&
\sum_{a_1...a_r}
{\cal C}_{a_1...a_r}
\sum_{k=1}^{N_r}
q^{\alpha_k^1}_{a_1}\cdots q^{\alpha_k^r}_{a_r}
f_k.
\non
\eea
After the second equal-sign we inserted the explicit form of the coefficients,
and use the fact that the color factor is independent of the shape of
the parallel transporter.
In the last step we combine the three summations into that over $k$.
More precisely, we did a relabelling of the indices,
and rewrote the amplitude factor.
Finally the resulting list of $P$, $L_k$ is created by putting
the new label structure and the new amplitude.
The algorithm is summarized as
\bi
\item Relabeling:
$\{\alpha_i^1,\cdots,\alpha_i^{s},\alpha_j^1,\cdots,\alpha_j^{r-s}\}\longrightarrow\{\alpha_k^1,\cdots,\alpha_k^{s},\alpha_k^{s+1},\cdots,\alpha_k^r\}$

\item Amplitude part:
$\frac{r!}{s!(r-s)!}f_i^{\prime}f_j^{\prime\prime}\longrightarrow f_k$

\item Creating list:
$L_i^{\prime}\times L_j^{\prime\prime}\longrightarrow L_k=(\alpha_k^1,\cdots,\alpha_k^{r};f_k)$
\ei
This procedure should be carried out for $0 \le s \le r$,
$1 \le i \le N_s^{\prime}$ and $1 \le j \le N_{r-s}^{\prime\prime}$
if order $r$ is desired.
The algorithm has been implemented in the python script language,
which is good at dealing with the complicated list operation.
In this way,
one can obtain the vertices for any parallel transporter.

\section{Extension to the Schr\"odinger Functional}
The essential new ingredient in the SF
is the presence of the non-vanishing back ground field.
It is induced by non trivial Dirichlet boundary conditions in  the
time direction.
In this case, the link variable is expressed by
the background field and the fluctuation field $q_{\mu}$,
\be
U(x,\mu)=V(x,\mu)e^{ g_0 q_{\mu}(x)}.
\ee
The main difficulty due to the presence of the back ground field
occurs in the color factor.
In the color factor,
the back ground field can appear  ``randomly'' between SU(3) generators, for example
\be
{\cal C}_{a_1 a_2 a_3 \cdots}=T_{a_1} T_{a_2} T_{a_3} \cdots
\longrightarrow
T_{a_1} V T_{a_2} V^{-1} T_{a_3} \cdots,
\hspace{5mm}
\mbox{on SF},
\ee
where arguments of $V$'s and powers ($V$ or $V^{-1}$)
depend on the location of the links and its orientation.
At first glance, it seems difficult to deal with it in a systematic way.
Furthermore, since the background field depends on
the Lorentz and position indices, the color factor
is located inside of the sum over $i$ and
does not  factorize anymore,
\be
P_r=
\sum_{a_1,...,a_r}
{\cal C}_{a_1\cdots a_r}
\sum_{k=1}^{N_r}
q^{\alpha_k^1}_{a_1}\cdots q^{\alpha_k^r}_{a_r}
f_k
\longrightarrow
\sum_{a_1,...,a_r}
\sum_{k=1}^{N_r}
{\cal C}^k_{a_1\cdots a_r}
q^{\alpha_k^1}_{a_1}\cdots q^{\alpha_k^r}_{a_r}
f_k,
\hspace{5mm}
\mbox{on SF}.
\ee

One can solve the problem however
for the restricted class of background fields
that have mainly been used in applications of the SF \cite{Luscher:1993gh},
$V(x,0)=1$, $V(x,k)=V(x_0)$.
In addition, we take the background field to be abelian,
given by diagonal color matrices.
This means that
the generators in the $V(x_0)$ are written as linear combinations of the
elements of the Cartan sub-algebra $H_j$,
that is, $V(x_0)=e^{i \sum_j h_j(x_0) H_j}$ with coefficients $h_j(x_0)$.
Now we find
the nice equation,
$V(x_0) I_a V^{-1}(x_0)=I_a e^{i\phi_a(x_0)}$, and
the $I_a$ (Cartan basis)
do not mix with other basis elements.
This is because, in the adjoint representation, an element of the Cartan basis
is an eigenstate of the Cartan generator,
$[H_j,I_a]=\mu_{ja}I_a$.
The eigenvalue $\mu$ is a root, and
the phase is a linear combination of the roots and $h_j(x_0)$,
$\phi_a(x_0)=\sum_j \mu_{ja}h_j(x_0)$.
Another property 
of the standard background fields that is relevant here
is that the back ground field and
the phase has a simple dependence on time,
$V(x_0+\Delta t)=V(x_0)\exp{(i \Delta t {\cal E})}$,
$\phi_a(x_0+\Delta t)=\phi_a(x_0)+ \Delta t \psi_a$,
where ${\cal E}$ is the color electric field.
This however needs to be relaxed later, see in next section.

Our main finding is
that by making use of the above properties of the background field
any color factor of order $r$ can be written as 
\be
{\cal C}^k_{a_1\cdots a_r}(x_0)
=
\frac{}{}
\underbrace{
\left[
I_{a_1} \cdots I_{a_r}
V^{A_k}(x_0)
e^{i {\cal E}B_k}
\right]
}_{3 \times 3 \mbox{ matrix}}
\underbrace{
e^{\frac{i}{2}\sum_{u=1}^r
(
\psi_{a_{u}}C_k^{(u)}
+
\phi_{a_{u}}(x_0)D_k^{(u)})}
}_{U(1) \mbox{ phase factor}},
\ee
where the background field has been moved
to the right in the $3\times 3$ matrix part.
Actually we can show by induction that
$A$ and $B$ are single component integer,
and $C$ and $D$ are $r$ component integer valued vectors.
In the expression,
the information of the lattice size
and the back ground field are encoded in
$V$, ${\cal E}$, $\phi$ and $\psi$.
The lists are independent of these values,
only $x_0,A,B,C,D$ are required.
The former are only needed when implementing the vertex
in a diagram calculation program at a second stage.
Note that we need $x_0$,
since there is no translation symmetry for the time direction.
We choose $x_0$ as the time component of the
position of the left most link variable in the parallel transporter.
The benefit of the expression is that
we can separate the information about the list ($A,B,C,D$)
and the lattice size and the back ground field.
Therefore, we can do a symbolic list operation,
independently of the details of the lattice and
the background field.

We have obtained a manageable expression for the color factor.
Next we formulate the multiplication for this structure of
integer lists $(x_0,A,B,C,D)$.
From an actual multiplication of the color factors,
we found the algorithm to get
a list of color factor ${\cal C}$ (order $r$)
from those of ${\cal C}^{\prime}$ (order $s$)
and ${\cal C}^{\prime\prime}$ (order $r-s$),
$(x_0,A,B,C,D)
\longleftarrow
(x_0^{\prime},A^{\prime},B^{\prime},C^{\prime},D^{\prime})
\times
(x_0^{\prime\prime},A^{\prime\prime},B^{\prime\prime},C^{\prime\prime},D^{\prime\prime}),
$
\bea
x_0
&\longleftarrow&
x_0^{\prime},
\\
A
&\longleftarrow&
A^{\prime}+A^{\prime\prime},
\\
B
&\longleftarrow&
B^{\prime}+B^{\prime\prime} + \Delta t A^{\prime\prime},
\\
C
&\longleftarrow&
(
\underbrace{
\underbrace{C_{(1)}^{\prime},\cdots,C_{(s)}^{\prime}}_{s
      \mbox{ elements}},
\underbrace{C_{(1)}^{\prime\prime}+2 B^{\prime}+\Delta
      t D_{(1)}^{\prime\prime},
\cdots,
C_{(r-s)}^{\prime\prime}+2 B^{\prime}+\Delta t D_{(r-s)}^{\prime\prime}
}_{r-s \mbox{ elements}}
}_{r \mbox{ elements}}
),
\\
D
&\longleftarrow&
(
\underbrace{
\underbrace{D_{(1)}^{\prime},\cdots,D_{(s)}^{\prime}}_{s \mbox{ elements}},
\underbrace{D_{(1)}^{\prime\prime}+2 A^{\prime},
\cdots,
D_{(r-s)}^{\prime\prime}+2 A^{\prime}}_{r-s \mbox{ elements}}
}_{r \mbox{ elements}}
),
\label{eqn:multABCD}
\eea
where $\Delta t=x_0^{\prime\prime}-x_0^{\prime}$.
It turns out that the resulting $A$ and $B$ remain single component integer.
On the other hand, the resulting $C$ and $D$ are given by
combinations of single prime and
double prime objects with some additional terms.
Since $A,B,C,D$ are all integer value
and this operation is simple,
the algorithm is suited for a symbolic operation and
easily implemented in python script language.
As a new ingredient in the implementation,
we have to add the new components $x_0,A,B,C,D$ to the earlier list structure,
\be
L_k^{(r)}=(\alpha_k^1,\cdots,\alpha_k^r,
x_0,A_k,B_k,C_k,D_k;f_k).
\ee
Even in the SF with non-trivial color factor,
the algorithm maintains a closed multiplication structure,
therefore it is applicable for any parallel transporter.

To confirm and check the algorithm,
we perform a one-loop calculation of the SF coupling.
By calculating this quantity,
we can check the two-point vertex, that is, inverse propagator.
We investigate not only the plaquette gauge action considered before
 \cite{Luscher:1993gh},
but also the improved gauge actions including the rectangular
loop \cite{Takeda:2003he} and
get consistent results.
Furthermore, we compared with
the hand derived three point vertex of the plaquette gauge action,
available from a private note of Peter Weisz
and confirmed consistency.
To get further confidence in the implementation of our algorithm,
we have to check the four-point vertex.
To do so we need to do a two-loop calculation of the SF coupling
\cite{Bode:1998hd,Bode:1999sm},
and this will be reported in the future.

\section{Application I: $L=T\pm a$ lattice}
\begin{figure}[t]
\begin{center}
 \begin{tabular}{cc}
  \scalebox{1}{\includegraphics{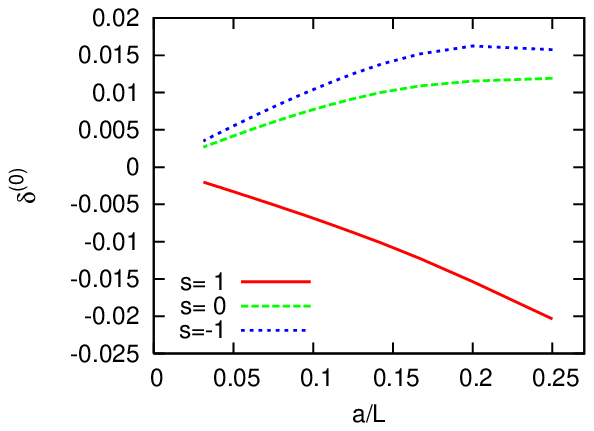}}&
  \scalebox{1}{\includegraphics{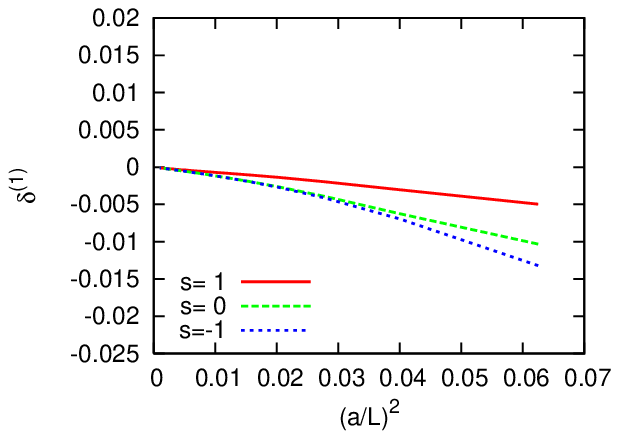}}\\
 \end{tabular}
\caption{One-loop relative deviation as a function of $a/L$.
The left(right) panel shows tree(one-loop) level $O(a)$ improved results
\label{fig:deviation}}
\end{center}
\end{figure}

As a simple novel application of our algorithm,
we perform a one-loop computation of the SF coupling
on lattices with $L=T+s a$ with $s=\pm 1$.
Such lattices are motivated by considering the SF with staggered fermions.
Due to the Dirichlet boundary in the time direction,
we have to set $T$ to be odd.
Here, we will discuss only the gauge part on the lattice, not
the staggered fermion part.
More details about the latter are given in the contribution of P.~Perez~Rubio
and S.~Sint
to this conference.

When one takes the continuum limit in the standard SF,
one sets the $T/L=1$.
The fact that we here have to set $(T+sa)/L=1$
when taking the continuum limit in the tree-level $O(a)$
improved theory
causes some change
to the solution of the equation of motion,
that is, the background field.
It is modified by
lattice artifacts as compared to the standard one,
and has to be extracted numerically from the equation of motion.
The time dependence of the background field turns out not to be
strictly linear anymore.
Therefore we have to extend our algorithm
to apply for an arbitrary time dependent phase $\phi(x_0)$.
For this purpose, we have derived another variant algorithm
for the color factor which is similar to eq.(\ref{eqn:multABCD}),
and it will be shown in more detail in a future publication.

In Figure \ref{fig:deviation},
we show the resulting relative deviation of the step scaling function to
one-loop order, $\delta^{(k)}_1(a/L)$, given by
\bea
\delta (u,a/L)
&=&
\frac{\Sigma(u,a/L)
 - \sigma(u)}{\sigma(u)}
=
\delta^{(k)}_1(a/L) u + O(u^2),
\\
\sigma(u)
&=&
\bar{g}^2(2L),
\hspace{5mm}
u=\bar{g}^2(L),
\eea
where $k=0$($k=1$) is the tree (one-loop)level O($a$) improved case.
In the plots, for comparison, we also include those of $s=0$
where $L$ and $T$ are the same.
As a result, we observe that all three cases have similar
absolute size of the lattice artifacts.

\section{Application II: $\Lambda$ parameter for improved gauge actions}
\begin{figure}[t]
\begin{center}
      \psfragscanon
      \psfrag{c0}[][][2]{$c_0$}
      \psfrag{c1}[][][2]{$c_1$}
      \psfrag{c2}[][][2]{$c_2$}
      \psfrag{c3}[][][2]{$c_3$}
      \scalebox{0.62}{\includegraphics{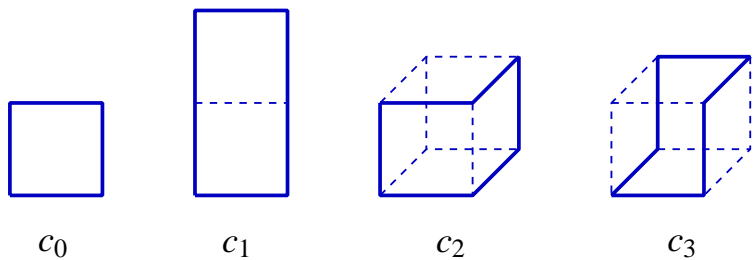}}
\caption{Loops with weight factors $c_{0,1,2,3}$
which contribute to the improved gauge action.
\label{fig:loop}}
\end{center}
\end{figure}

As a next application,
we apply the algorithm to the improved gauge actions
including six link loops (not only for the
rectangular type but also for
the chair and 3-dimensional type actions).
The loops are shown in Figure \ref{fig:loop} with
weight factors, $c_0$, $c_1$ etc. and the weights are normalized
by $c_0+8c_1+16c_2+8c_3=1$.
We perform the one-loop computation
of the SF coupling for the improved gauge actions
and extract information about
a ratio of lambda parameters for the pure SU(3) gauge theory
between the
lattice scheme (the various gauge actions)
and the SF scheme, $\Lambda_{\rm Lat}/\Lambda_{\rm SF}$.
By combining the result of  $\Lambda_{\rm SF}/\Lambda_{\overline{\rm MS}}$ in
\cite{Bode:1999sm}
for $N=3$ and $N_{\rm f}=0$,
we summarize the values of
$\Lambda_{\rm Lat}/\Lambda_{\overline{\rm MS}}=
\Lambda_{\rm Lat}/\Lambda_{\rm SF}\cdot
\Lambda_{\rm SF}/\Lambda_{\overline{\rm MS}}$
in Table \ref{tab:ratios}.
We observe rough consistency with old results.

\begin{table}[t]
 \begin{center} 
  \begin{tabular}{|c|lll|l|}
  \hline\hline
  Action
& \multicolumn{1}{c} {$c_1$}
& \multicolumn{1}{c} {$c_2$} 
& \multicolumn{1}{c|}{$c_3$} 
& $\Lambda_{\rm Lat}/\Lambda_{\overline{\rm MS}}$
\\ \hline
Plaquette  &$\phm 0$  &$\phm 0$   &$\phm 0$  &$\pho 0.0347109675...$
\cite{Luscher:1995nr}\\ \hline
Wilson RG  &$-0.252$  &$\phm 0$   &$-0.170$  &$\pho 2.34086 (2 )$ \\ \hline
Iwasaki    &$-0.331$  &$\phm 0$   &$\phm 0$  &$\pho 2.12455(5)$ \\ \hline
DBW2       &$-1.40686$&$\phm 0$   &$\phm 0$  &$44.21 (2)$ \\ \hline
Symanzik   &$-1/12$   &$\phm 0$   &$\phm 0$  &$\pho 0.1836938(4)$ \\ \hline
Symanzik II&$-1/12$   &$\phm 1/16$&$-1/16$   &$\pho 0.1782883(4)$ \\ \hline
no name    &$-1/12$   &$-0.1$     &$\phm 0.1$&$\pho 0.1673674(8)$ \\
  \hline\hline
  \end{tabular}
\caption{We show our results of
$\Lambda_{\rm Lat}/\Lambda_{\overline{\rm MS}}$
in SU(3) gauge theory for various gauge actions,
except for the plaquette gauge action
ref. \cite{Luscher:1995nr} which is given here for completeness.
Our values are roughly consistent with
old results \cite{Weisz:1983bn,
Iwasaki:1983zm,Iwasaki:1984cj,Ukawa:1983ae,Bernreuther:1984wx,Sakai:2000jm},
but in ref.\cite{Sakai:2000jm}
the authors do not show error estimates therefore,
it is hard to compare.
After the completion of this work, new results for
several gauge actions have appeared in ref. \cite{new}.
\label{tab:ratios}}
 \end{center}
\end{table}

We thank the Deutsche Forschungsgemeinschaft (DFG)
for support in the framework of SFB Transregio 9.


\begin{thebibliography}{99}
\bibitem{Luscher:1985wf}
M.~L\"uscher and P.~Weisz,
\newblock Nucl. Phys. {\bf B266} (1986) 309.

\bibitem{Hart:2004bd}
A.~Hart, G.~M. von Hippel, R.~R. Horgan and L.~C. Storoni,
\newblock J. Comput. Phys. {\bf 209} (2005) 340.

\bibitem{Luscher:1993gh}
M.~L\"uscher, R.~Sommer, P.~Weisz and U.~Wolff,
\newblock Nucl. Phys. {\bf B413} (1994) 481.

\bibitem{Takeda:2003he}
S.~Takeda, S.~Aoki and K.~Ide,
\newblock Phys. Rev. {\bf D68} (2003) 014505.

\bibitem{Bode:1998hd}
A.~Bode, U.~Wolff and P.~Weisz,
\newblock Nucl. Phys. {\bf B540} (1999) 491.

\bibitem{Bode:1999sm}
A.~Bode, P.~Weisz and U.~Wolff,
\newblock Nucl. Phys. {\bf B576} (2000) 517.

\bibitem{Luscher:1995nr}
M.~L\"uscher and P.~Weisz,
\newblock Phys. Lett. {\bf B349} (1995) 165.

\bibitem{Weisz:1983bn}
P.~Weisz and R.~Wohlert,
\newblock Nucl. Phys. {\bf B236} (1984) 397.

\bibitem{Iwasaki:1983zm}
Y.~Iwasaki and S.~Sakai,
\newblock Nucl. Phys. {\bf B248} (1984) 441.

\bibitem{Iwasaki:1984cj}
Y.~Iwasaki and T.~Yoshie,
\newblock Phys. Lett. {\bf B143} (1984) 449.

\bibitem{Ukawa:1983ae}
A.~Ukawa and S.-K. Yang,
\newblock Phys. Lett. {\bf B137} (1984) 201.

\bibitem{Bernreuther:1984wx}
W.~Bernreuther, W.~Wetzel and R.~Wohlert,
\newblock Phys. Lett. {\bf B142} (1984) 407.

\bibitem{Sakai:2000jm}
S.~Sakai, T.~Saito and A.~Nakamura,
\newblock Nucl. Phys. {\bf B584} (2000) 528.

\bibitem{new}
A.~Skouroupathis and H.~Panagopoulos,
{\tt hep-lat/0709.3239}


\end{thebibliography}

\end{document}